\documentclass{ptapap}

\author{Agnieszka Janiuk}[CFT]
\author{Szymon Charzy{\'n}ski}[FUW, CFT]
\affil[CFT]{Center for Theoretical Physics\\
  Al. Lotnikow 32/46, 02--668 Warszawa, Poland}
\affil[FUW]{Chair of Mathematical Methods in Physics, University of Warsaw\\ ul. Pasteura 5, 02-093 Warszawa, Poland}

\title{Simulations of coalescing black holes}

\begin{document}

\maketitle

\begin{abstract}

We describe the methods and results of numerical simulations of coalescing black holes. The simulation in dynamical spacetime covers the inspiral, merger, and ringdown phases. We analyze the emission of gravitational waves and properties of a black hole being the merger product. We discuss the results in the context of astrophysical environment of black holes that exist in the Universe.
\end{abstract}

\section{Introduction}

\subsection{Binary black holes in the Universe}

Binary black holes can be quite ubiquitous in the Universe. The massive stars are expected to form BHs at the end of their lives. We already know a few dozen of stellar mass BHs that are in X-ray binary systems, so that in case of a high mass X-ray binary, it is expected that the system evolves to form of a binary black hole.
Supermassive black holes ($10^{6}-10^{8} M_{\odot}$) are thought to exist in the centers of most galaxies. Galaxy merging process observed to be ongoing
in number of cases \citep{2015Volonteri} inevitably involves the merger of a binary black hole as the final phase of galaxy merger.
One example is the pair of merging galaxies in the cluster
Abell 400,
also, the kinked jet, observed in the galaxy NGC 326, suggests that the merger process might have happened there in the past.


\subsection{Two body problem}

In Newtonian gravity, the solution of the two body problem is a bound orbit. The motion of a planet occurs along an ellipse, focused onto the Sun.
The two body problem in general relativity cannot aim for an analytical solution.
This is because the gravitational field carries energy, momentum, and angular momentum, so, in some sense it is the third physical object involved in the problem.
For large enough separation of components, the post-Newtonian approximation works. In this case, the components of a binary system fall onto each other. The energy and angular momentum is radiated away through the gravitational waves, but in this regime the gravitational radiation is treated as a small perturbation to the Newtonian motion.

In case of a strong gravitational field, fast motion of the binary components and small separations, the problem may be solved only numerically.
However, for many years, the attempts to numerically solve the strong field regime have failed. The perturbation theory was used to describe the final phase of a 'ringdown', after the black holes had already merged.
Also, the test particle motion could be described using approximate techniques,
to follow the geodesic motion in the vicinity of a black hole. The effects such as the perihelion precession, or the unstable and chaotic orbits, were studied.
Nevertheless, the full numerical treatment of the
binary black hole system evolution is extremely difficult.
If we need to follow the completely different phases, i.e., the inspiral, merger, and ringdown, the main problems to overcome are
the numerical instabilities of discretized Einstein equations.

The field equations of general relativity
$$
G_{\mu\nu} = 8 \pi T_{\mu\nu}
$$
written in terms of the spacetime metrics
$ds^{2} = g_{\mu\nu}dx^{\mu}dx^{\nu}$,
form a system of 10 coupled, non-linear, second order, partial differential equations. Each of them depends on the 4 space-time coordinates.
For a numerical scheme, one must choose the coordinates and set of evolved variables, in a way providing numerical stability of the simulation. For the constrained evolution, the system of hyperbolic and elliptic equations must be solved. The numerical scheme must also deal with geometric singularities in BH spacetimes.
Another source of difficulties is the fact, that BH merger is a many scale problem.
On one hand, the strong field region around BH's has to be simulated with a high resolution, but
on the other hand, a large volume has to be covered by the simulation to include the radiation effects.

The first successful simulation of BBH merger dates to the year 2005.
The BH singularities problems are solved by excision algorithms or by the moving puncture methods.
Also, the novel technique of adaptive mesh refinement of the simulation grid is used to cover large volume with low resolution and to cover regions around black holes with a high resolution grid.
Typically, in such simulations up to 10 levels of refinement are used (grid spacing differs by a factor of $2^{10}$).

The BSSN method (Baumgarte-Shapiro-Shibata-Nakamura) proved to be the most stable numerical formulation of 3+1 decomposition of Einstein equations and is used for discretization of evolution equations of spacetime geometry.
The BSSN formalism is a modification of the ADM formalism developed during the 1950s. For more complete description of the methods, we refer the reader to \citet{2005CQGra..22..425P,2006PhRvL..96k1101C,2006ApJ...653L..93B}.


\subsection{BSSN formulation}

In addition to usual 3+1 decomposition of the spacetime metric
$$
ds^{2} = g_{\mu \nu} dx^{\mu} dx^{\nu} = -\alpha^{2} dt^{2} + \gamma_{ij} (dx^{i}+\beta^{i} dt)(dx^{j}+\beta^{j}dt)
$$
the 3-metric $\gamma_{ij}$ is conformally transformed via
$$
\phi=\frac{1}{12}\ln\det\gamma_{ij}, \qquad \tilde{\gamma}_{ij}=e^{-4\phi}\gamma_{ij}
$$
so $\det\tilde{\gamma}_{ij}=1$.
The extrinsic curvature $K_{ij}$ is also transformed:
$$
K:=\mathop{tr} K_{ij}=g^{ij}K_{ij}, \qquad \tilde{A}_{ij}=e^{-4\phi}\left(K_{ij}-\frac13\gamma_{ij}K\right)
$$
with $\mathop{tr}\tilde{A}_{ij}=0$.\\
The evolved variables are: $\phi$, $\tilde{\gamma}_{ij}$, $K$, $\tilde{A}_{ij}$ and $\tilde{\Gamma}^i:=\tilde{\gamma}^{jk}\tilde{\Gamma}^i_{jk}$.

\begin{figure}
  \centering
  \begin{minipage}{0.48\textwidth}
    \includegraphics[width=\textwidth]{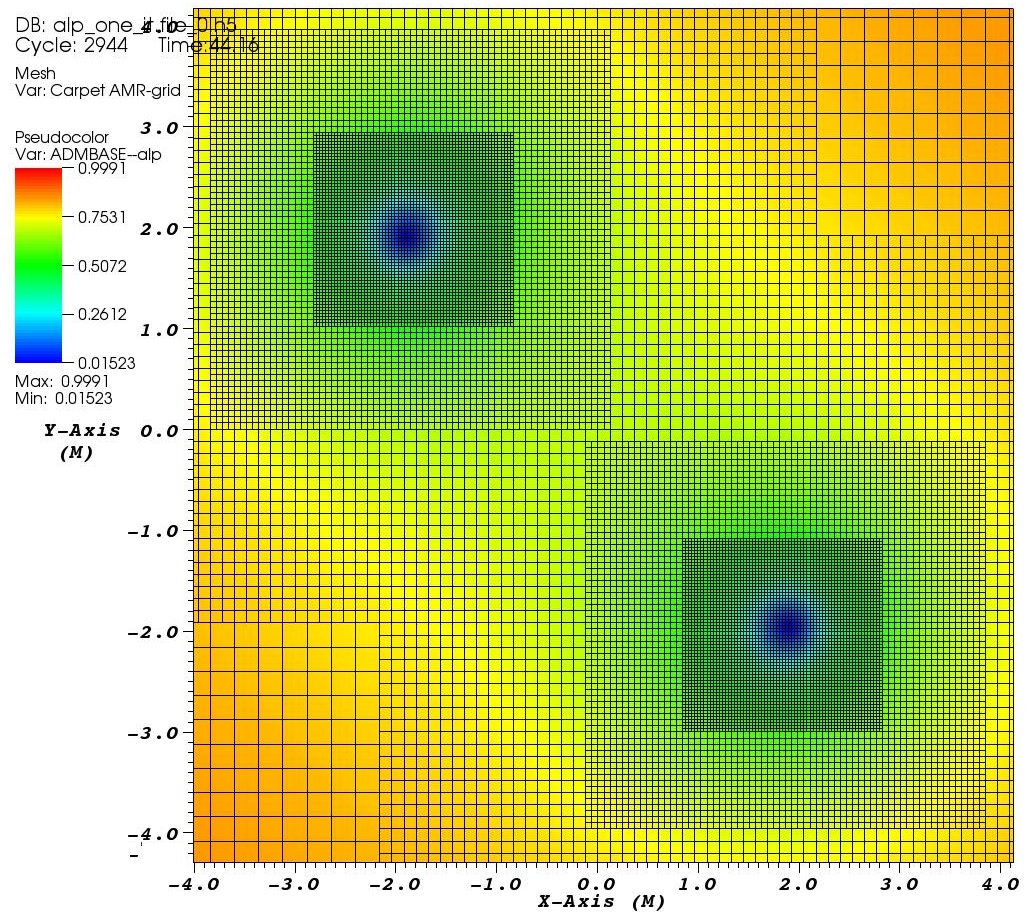}
    \vspace{-25pt}
    \label{fig:face1}
    \caption{Adaptive mesh refinement in the simulation of merging black holes.}
  \end{minipage}
  \quad
  \begin{minipage}{0.48\textwidth}
    \includegraphics[width=\textwidth]{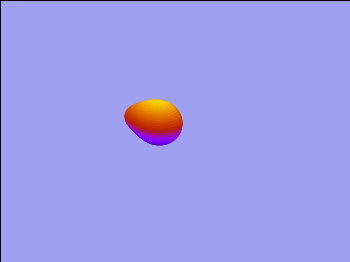}
    \label{fig:face2}
    \caption{Shape of an instantaneous apparent horizon of the pair of merged
black holes.}
  \end{minipage}
\end{figure}

\subsection{Gravitational waves analysis and recoil}

Analyzing the properties of spactime after the merger one can easily compute the mass and spin of the black hole being the final state. However, the estimation of its recoil velocity is a more subtle problem. It requires the evaluation of the linear momentum carried away by the gravitational radiation during the merger. To calculate the total momentum, we follow the algorithm described by \citet{2008itnr.book.....A}.
The formula for $d\vec{P}/dt$ is expressed in terms of the coefficients $A^{\rm lm}$ of multipole expansion of the Weyl scalar $\psi_4$.
Then,  the coefficients $A^{\rm lm}$ are computed
on the sphere of large radius (as large as the simulated volume), with $l$ ranging from 2 to 4.  We integrate $d\vec{P}/dt$ over the time of the simulation.
Since the total momentum has to be conserved, we are able to compute the recoil of the merged black hole.
The problem of gravitational recoil is widely studied. For properly chosen mass ratios and spin configurations of the initial BHs, the kick velocity can exceed 4000 km/s, but such large recoils are very rare. Resent statistical estimations \citep{2011PhRvL.107w1102L} of the kick velocity probabilities are presented in the Table below.

\smallskip

\noindent
\begin{minipage}{0.55\textwidth}
The probability $P_{\mathrm{obs}}$
takes into account that one of the main
methods to search for recoiling black holes is to look for a
large differential redshift. Therefore, it
refers to the values of recoil velocity projected along the
line of sight.
\end{minipage}
\quad
\begin{minipage}{0.4\textwidth}
\small
\noindent
\begin{tabular}{|l|c|c|}
\hline
range [km/s] & $P$ & $P_{\mathrm{obs}}$\\
\hline
0-500 & 79.027\%  & 92.641\%\\
500-1000 & 15.399\%  & 6.177\%\\
1000-2000 &  5.384\%  & 1.164\%\\
2000-3000 &  0.189\%  & 0.018\%\\
3000-4000 &  0.001\%  & 0.0001\%\\
\hline
\end{tabular}
\end{minipage}


\section{BBHs and Gamma ray bursts}

Gamma Ray Bursts (GRBs) are the extremely energetic, transient events observed isotropically in the sky in the high energy bands.
Their known cosmological origin requires the physical process that produces them, to be the cosmic explosion of an enormous power.
The two possible mechanisms involve the birth of a new black hole in a
cataclysmic event, one of which is connected with the collapse of a massive star, or the merger of two compact objects, binary neutron stars, or BH-NS binary.
They should be responsible for long (>2 seconds) or short GRBs, respectively.
The so called 'central engine' of such process is in both cases the hot and
dense accretion disc with a hyper-Eddington accretion rate (up to a few $M_\odot s^{-1}$) around a spinning black hole, which triggers the powerful, ultra-relativistic jets. These jets produce gamma ray radiation far away from the central region.


In the new proposed process we combine the two above scenarios, namely the collapse of a massive star, with a binary black hole merger.
The progenitor, i.e. the collapse of a massive rotating star, should
occur in a close binary
system with a companion BH. A candidate for such scenario is the future of Cygnus X-3,
which is the X-ray binary composed
of a Wolf-Rayet star and black hole of a mass about $3 M_{\odot}$ \citep{2013MNRAS.429L.104Z}. If the companion enters the outer part of the star's envelope, the tidal squeezing effect may help trigger the collapse of the star's core into a black hole \citep{1985MNRAS.212...57L}. This one should be at least moderately, or fast rotating, as the envelope will be additionally spun up by
the transfer of the angular momentum from the companion.
We envisaged such scenario as a new
 model for a GRB, whose extremely long duration (> 1000 s) would be associated with the presence of
two jets \citep{2013A&A...560A..25J}. In some cases, one or both jets can be redirected from the line of sight \citep{2016arXiv160407132J}. Also, jet precession is possible, or
the GRB engine leaving an 'orphan' afterglow, due to the gravitational kick.
In any case, the observed GRB should be
accompanied by a gravitational wave signal.

Our computations involve 3 stages of the system evolution, namely (i) the core collapse and accretion
onto the core, (ii) evacuation of
matter from the innermost region and binary black hole merger inside an outer circumbinary disk,
 and (iii) further accretion of the remaining matter onto the merged black hole.
Stages (i) and (iii) can be the engines of jets launching, and the resulting gamma rays; stage (ii) will be the engine of the gravitational wave. In addition, the merged black hole will receive the kick velocity,
depending on the BH mass ratio and their spin vectors configuration.


  \begin{minipage}{0.22\textwidth}
\vspace{5pt}
    \includegraphics[width=\textwidth]{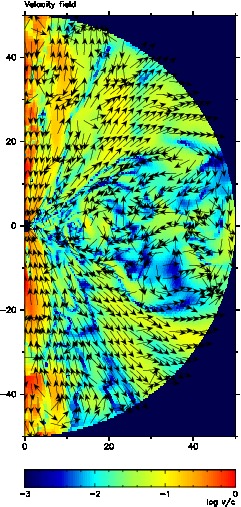}
    \label{fig:velfield}
  \end{minipage}
\hspace{0.03\textwidth}
  \begin{minipage}{0.70\textwidth}
The first stage, core-collapse, is computed semi-analytically.
We consider either a homologous accretion of the envelope and substantial
increase of the core (primary) BH mass, or the accretion through a torus, and wind outflow, which is supported by a centrifugal force and driven magnetically.
It can take away up to about 75\% of the torus mass,
so that the increase of the primary black hole mass is
moderate. This (rough) estimate is based on
the results given by \citet{2013ApJ...776..105J}.
The Figure shows the velocity field in the torus and wind, taken from a snapshot of GR MHD simulation of accretion onto a Kerr black hole. Arrows are normalized to unity, and the magnitude of velocity with respect to the speed of light is
depicted by the colors.
  \end{minipage}

The second phase of the event, which is
the evolution of binary BH
system, when the separation of its components becomes so small, that the full set of Einstein
equations is solved numerically,
is computed separately.
We consider two
black holes in a quasi circular orbit. An exemplary computation assumes the
mass ratios varying from 1 to 3. The
more massive black hole carries also spin perpendicular to the orbital plane,
and the second component is spinless.
In the first approximation, we neglect the influence of the surrounding matter (circum binary disc) on the BBH merger and simulate dynamics of the vacuum solution.
Note that the parameters in the Table are evaluated in geometric units. Since the vacuum solutions of Einstein equations are scalable they can be easily rescaled to fit the astrophysical constrains given above.

\begin{center}
{\scriptsize
\setlength\tabcolsep{1pt}
\hspace{-3pt}\begin{tabular}{|c|c|c|c|c|c||c|c|c|c|c||c|c|}
 \hline
 & \multicolumn{10}{|c||}{ Initial state} & \multicolumn{2}{|c|}{Final state}
\\
 \hline
 & \multicolumn{5}{|c||}{Control parameters}
 & \multicolumn{5}{|c||}{ Computed ADM values}
 & \multicolumn{2}{|c|}{ADM}
\\
 \hline
 run & $m_1$ & $m_2$ & $p_1$ & $p_2$ & $a_1$ & $M_1$ & $M_2$ & $\frac{M_1}{M_2}$ & $M$ & $\frac{a_1}{M_1^2}$ & $M_3$ & $\frac{a_3}{M_3^2}$
\\
 \hline
 A & 0.63 & 0.32 & -0.17 & 0.17 & 0.9 & 1.05 & 0.35 & 3.0 & 1.39 & 0.81 & 1.34
 & 0.76 
\\
 \hline
 B & 0.54 & 0.45 & -0.14 & 0.14 & 0.28 & 0.6 & 0.45 & 1.4 & 1.03 & 0.79 & 0.98
 & 0.78 
\\
 \hline
\end{tabular}
}
\end{center}

These simulations were performed using \texttt{Einstein Toolkit} computational package ({\tt http://www.einsteintoolkit.org}). It is a family of codes  for use in relativistic astrophysics, based on finite difference computation on a gridded mesh \citep{2012CQGra..29k5001L}. The Toolkit is supported by a distributed model, combining core support of software, tools, and documentation in its SVN and GIT repository with partnerships of developers.
We use the source code which we configured, compiled and run in the computer cluster
in the Warsaw University (ICM).
Our simulation initial data contain the given masses, momenta
and spins for each BH.  The initial separation of components is
equal to $6 M$ ($M$ is approximately equal to $ADM$ mass of the whole system).
Then, the evolution of the system is followed,
using
the BSSN method.
We adopt the Cartesian grid with
$48\times48\times48 M$, and resolution of $\Delta x=1.6M$,
and 7 levels of the adaptive mesh refinement by factor 2 in two regions around singularities.

The apparent horizons are localized
around the components of the BH system and around final merged black
hole after it forms. The proper integrals over the isolated horizons
are calculated
to extract the values of mass and spin of the binary components and the
merger product. The values of recoil speeds obtained from analysis of gravitational radiation for the cases presented in the table are approximately 200~km/s and 300 km/s (for the runs $A$ and $B$, respectively).

\begin{figure}
  \centering
  \begin{minipage}{0.48\textwidth}
   \centering
    \includegraphics[width=0.75\textwidth]{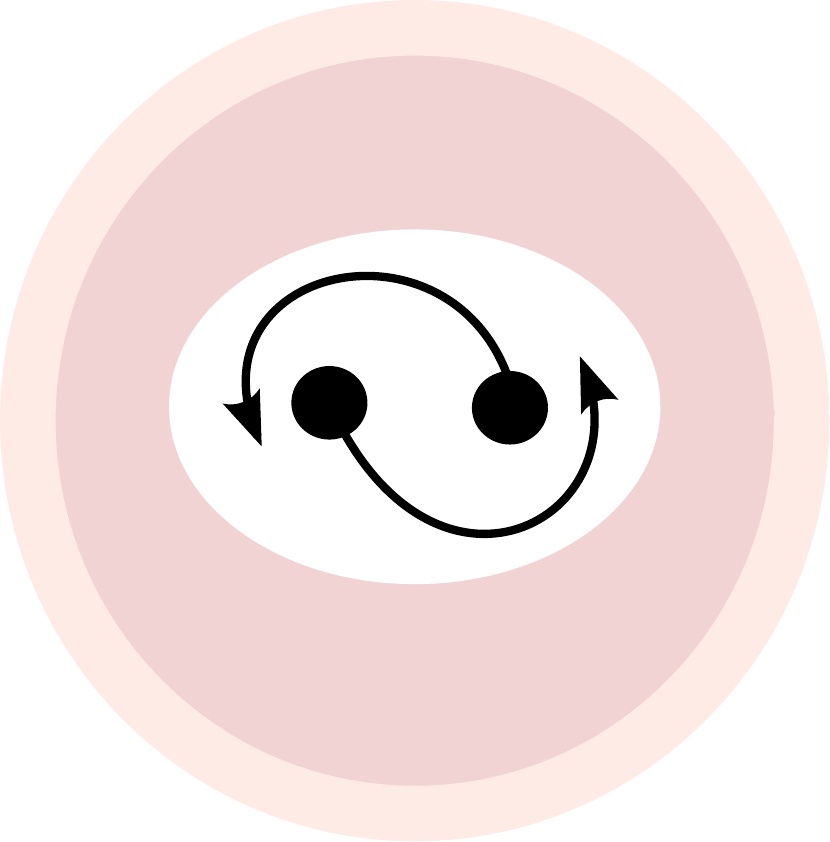}
    \label{fig:face1}
    \caption{Binary black holes merging in a quasi-vacuum, well inside a circumbinary disk.}
  \end{minipage}
  \quad
  \begin{minipage}{0.48\textwidth}
    \centering
    \includegraphics[width=\textwidth]{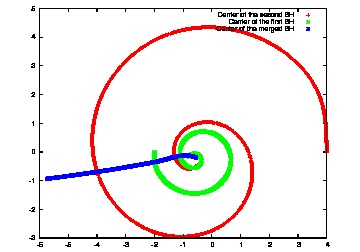}
    \label{fig:face2}
    \caption{Example trajectories of the non-equal mass coalescing black holes, and their merger product.}
  \end{minipage}
\end{figure}



\section{Summary and discussion}

Regardless of the above scenario, the core of massive star can begin to collapse
because of the tidal squeezing induced by the companion black hole.
This effect was invoked in the context of formation of the Thorne-Zytkov objects \citep{2012ApJ...752L...2C}
and has been discussed also in case of stars that are
tidally disrupted by
the supermassive black holes \citep{2013MNRAS.435.1809S}. In general, a significant orbital eccentricity is needed to induce a strong tidal compression.
Therefore, a perturbation from a third body, or other way of synchronizing
the inspiral of the secondary black hole with the core collapse
of the primary, would more efficiently trigger the event.
Also, in case of a moderately strong outflow from the accretion disk,
its outer parts may still be large enough to exhibit some misalignment from the
plane perpendicular to the black hole spin vector. In this case, the disc and
jet precession may occur.


Numerical relativity for the past several years has brought novel methods to discretize the field equations and follow the black hole merger process and its consequences. On the other hand, the observational evidences for spin flips and past merger events have been hinted, so that the theoretical effort is supported and motivated by observational discoveries.
In this work, we discussed the numerical methods used to study the binary black hole coalescence problem. In particular, we presented our studies of a possible 'exotic' gamma ray burst progenitor.
Obviously, the fully consistent treatment of the problem should involve the
 environmental issues (MHD, radiation) and this is the next challenge after the vacuum process is tackled.

\acknowledgements{We thank Petra Sukova and
Michal Bejger for helpful discussions.
We also acknowledge support from the
Polish National Science Center (DEC-2012/05/E/ST9/03914).}

\bibliographystyle{ptapap}
\bibliography{ptapap_ajaniuk}

\end{document}